\documentstyle[twocolumn,aps]{revtex}
\begin{document}
\input epsf
\draft
\title{\bf Implications for the Cosmic Ray Spectrum of a Negative Electron
Neutrino (Mass)$^2$}
\author{Robert Ehrlich}
\address{Physics Department, George Mason
University, Fairfax, VA 22030}
\date{\today}
\maketitle

\begin{abstract}

The features and problems of a speculative model based on the electron 
neutrino being a tachyon are discussed.  The model is consistent with five
properties of the cosmic ray spectrum, and it predicts a flux of neutrons
in a narrow energy region centered on $4.5\pm 2.2\times 10^{15} eV.$

\end{abstract}

%
%

%
%

%


\section{Introduction}

Following a suggestion by Kosteleck\'{y},\cite{Kostelecky} we posit the 
electron neutrino to be a tachyon with $|m_{\nu}|\equiv\sqrt{-m^2} 
\approx 0.5 eV/c^2,$ and consider the consequences for the cosmic ray 
(CR) spectrum.  The hypothesis, while it is highly speculative, is 
consistent with other neutrino observations, and it predicts the 
existence of a CR neutron flux in a narrow range of energies centered on 
$4.5\pm 2.2\times 10^{15} eV$.

Tachyons, first postulated in 1962, by Bilaniuk, Deshpande, and 
Sudarshan,\cite{Bilaniuk} are taken seriously by few physicists, because 
of the paradoxes they create, and because nearly\cite{Clay} all 
experiments specifically\cite{tritium} searching for tachyons have 
turned up negative.\cite{Alvager,Baltay}  Whatever one's view of 
tachyons, their existence is clearly an experimental question.  Weakly 
interacting tachyons of low mass would have probably escaped detection, 
or else not be recognized as such.  In fact, Chodos, Hauser and 
Kosteleck\'{y} suggested in 1985 that neutrinos are 
tachyons.\cite{Chodos85}

Chodos et al.\cite{Chodos92,Chodos94} noted that one could test
this hypothesis using a strange tachyon property, i.e., that particle 
decays producing tachyons which are energetically forbidden in one 
reference frame are allowed in another.  Thus, consider the ``decay": 
$p\rightarrow n+e^{+}+\nu_e$. For the decay to conserve energy in the 
proton rest frame, we need $E_\nu < 0$.  Now, tachyons, unlike
other particles, have $E<p$ so they can change the sign of $E$
when boosted to a sufficient velocity.  Thus, the tachyon energy in the 
proton rest frame, $E_\nu$ has the opposite sign from its energy in the lab
$E_{lab} = \gamma(E_\nu + \beta p_\nu\cos\theta)$ when $\beta$ exceeds
$-E_\nu/p_\nu\cos\theta < 1$.

The threshold lab energy for protons to decay is
found by making $E_{\nu}$ the least negative it can be in the CM frame, 
i.e., $-E_{\nu} = m_n + m_e - m_p \equiv \Delta$, and taking
$\cos\theta = 1.$  Therefore, at threshold
$\beta_{th}= -E_{\nu}/p_{\nu} \approx 1 + \frac{1}{2} 
m_{\nu}^2/{E_{\nu}}^2$, and hence
$\gamma_{th}=(1-\beta_{th}^2)^{-1/2}=\Delta/|m_{\nu}|$, so that

\begin{equation}
 E_{th} = \gamma_{th}m_p = \frac{m_p\Delta}{|{m_{\nu}}_e|}=
\frac{1.7\times10^{15}}{{|m_{\nu}}_e|}eV
\end{equation}

For nuclei of mass number A, $m_p$ is the mass of the parent nucleus, 
and $\Delta = m(A,Z\pm 1) + m_e - m(A,Z)$.  The idea of ``stable" 
particles decaying is less paradoxical if one reinterprets the emitted 
$\nu$ with $E_\nu > 0$ in the lab frame to be an absorbed $\bar{\nu}$ 
with $E_{\bar{\nu}} < 0$ from a background sea in the proton rest frame 
-- the so-called ``reinterpretation principle."\cite{Bilaniuk,Chodos85}  
This antineutrino background sea defines an absolute reference frame, presumably
coincident with that defined by the cosmic background radiation (CBR).

In order to test the prediction of Chodos et al. as applied to the CR 
spectrum, we need to calculate the mean free path for protons and other 
stable cosmic ray nuclei to decay as a function of their energy.  
Although we can easily deduce the threshold for such decays from 
kinematic arguments, finding the decay rates requires a knowledge of 
tachyon dynamics.  One might assume that the phase space involving 
negative energy tachyons could be treated in a similar manner as for 
positive energy particles.  Under this assumption the decay rate for 
$p\rightarrow n+e^++\nu$ could be estimated by integrating that tiny 
region of phase space in the CM for which $E_{\nu}$ changes sign between 
the CM and lab frames, and also assuming that the usual weak interaction 
coupling constant applies to the process.  However, the validity of such 
an approach is questionable.  Given the reinterpretation
principle, the rates for the processes $p\rightarrow n+e^++\nu$ and 
$\bar{\nu}_{bs}+ p\rightarrow n+e^+$ must be identical for any given
proton energy.  But the reaction rate of the latter reaction depends on 
both known antineutrino cross sections as well as the unknown density of 
antineutrinos in the background sea $(\bar{\nu}_{bs})$, and hence we 
have no way to estimate reliably the $p\rightarrow n+e^++\nu$ decay 
rate.  This being the case, we simply make an assumption that holds 
promise for explaining the knee of the CR spectrum: at all proton energies
significantly above threshold that the rate for proton decay greatly exceeds
that for conventional neutron decay.

\section{Modelling the Cosmic Ray Spectrum}

The idea that tachyonic neutrinos might explain the knee of the CR 
spectrum was first raised by Kosteleck\'{y}, though he regarded the 
existence of the knee by itself as insufficient evidence for the 
hypothesis in view of other explanations of the knee.\cite{Kostelecky}  
Moreover, Kosteleck\'{y} neither modelled the CR spectrum, as is done 
here, nor mentioned the signature neutron spike.  The inputs to the 
model are assumptions for: (1) $|m_{\nu_e}|$ values, (2) the energy 
spectrum and composition of CR's at their source, and (3) the spatial 
distribution of sources.

For the spatial distribution, we take an admixture of ``near" and ``far" 
sources.  Near sources are assumed to create CR's having path
distances to Earth from $10^4$ to $2\times 10^6$ ly, and far
sources are assumed to have path distances from
$2\times 10^6$ to $10^8$ ly.  For the source spectrum we use an 
$E^{-2.67}$ power law that fits the spectrum up to $10^{15} eV.$  
Essentially, we assume that the source spectrum is $E^{-2.67}$ for all 
$E$, and that changes in the observed spectrum are due to
particles in a given energy bin being shifted to lower energies as a 
result of beta decay.  Since the composition of CR's
above the knee is not well known, we try various compositions to
fit the data.  

The Monte Carlo method was used to obtain Figs. 1-3.  Protons
and nuclei were generated at various distances from Earth, and the fate 
of all particles in a given energy bin was considered to be the same, as 
their progress toward Earth was followed. For protons leaving sources 
above the threshold energy for decay, there is a chain of decays 
$p\rightarrow n\rightarrow p\rightarrow n\rightarrow p\cdots$ which 
stops when the nucleon either reaches Earth or else has its energy 
reduced below threshold.  As long as $E$ is above threshold, the
nucleon spends most of its time en route from the source as a 
rectilinearly propagating neutron, because the mean free path for 
neutrons before they decay is much greater than that for protons except 
quite close to $E_{th}.$  A similar decay chain occurs in
the case of $A > 1$ CR nuclei.
After each decay the daughter nucleus
has less energy in the lab frame than the parent.
Calculating the energy 
loss of the nucleon in a conventional beta
decay such as $n\rightarrow p+e^{-}+ \bar{\nu_e}$ is straightforward. In 
the CM frame 

\begin{figure}[htb]
\begin{center}
\leavevmode
\epsfxsize=3.25in
\epsffile{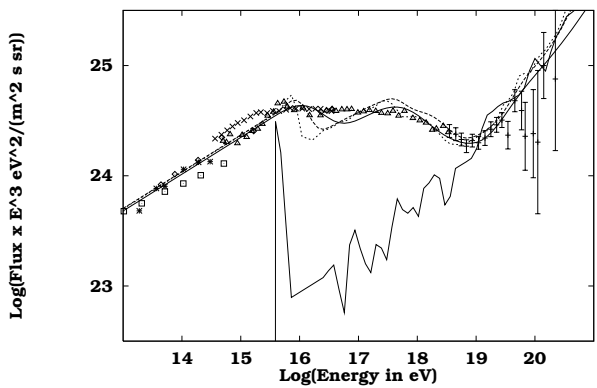}
\caption{Upper solid curve shows the prediction of the model for the CR 
flux, dn/dE, $(\times E^{3})$ assuming a tachyon mass $|m| = 0.5 
eV/c^2$, with convolution, using an energy resolution of $\Delta log 
E=\pm 0.4$.  The two dashed curves show fits with $|m_\nu| = 0.25 
eV/c^2$: short dashed curves assumes $\Delta log E= 0$, and the long 
dashed curve uses $\Delta log E=\pm 0.2$. The lower solid curve shows 
the predicted neutron spectrum component using $|m| = 0.5 eV/c^2$ and 
$\Delta log E= 0$.  All curves assume 13 \% near sources with mass 
compositions noted in the text.  Points are the data from: JAYCEE 
(diamonds), AGASA (with error bars), Aoyama-Hirosaki (squares), Tibet 
(crosses), Akeno 1km$^2$ array (diamonds), Proton Satellite 
(asterisks).} \end{center} \end{figure}

the proton has very little energy following the decay, and 
hence in the lab frame the nucleon loses a constant fraction $f\approx 
(1 - m_p/m_n)$ of its energy.  For the energetically forbidden decay, 
such as $p\rightarrow n+e^{+}+ \nu_e,$ the situation is more complex.
Here for proton lab energies much above threshold the neutrino needs to 
have highly negative energies in CM so that its energy in the lab frame 
be positive, and hence the daughter nucleus energy can no longer be 
ignored in the CM frame.  The calculation can be done as a sequence of 
two two-body decays: e.g., $p\rightarrow m(n,e^{+})+ \nu_e$ followed by 
$ m(n,e^{+}) \rightarrow n + e^+$, where in the first decay we choose 
only those events having $E_{lab} > 0$.

We show in Fig. 1 the log of the all particle flux $(\times E^3)$ -- 
both data and calculation.  A reasonably good fit to the spectrum is 
obtained for $|m_{\nu}|= 0.5 eV/c^2$ (solid curve), assuming that 13 \% 
of sources are ``near," with elemental abundances: 70\% A=1, 10\% A=4, 
10\% A= 5 to 19, 5\% A = 20 to 40, and 5\% A = 41 to 90.  The solid 
curve convolutes the Monte Carlo results with an energy resolution 
$\Delta log E = \pm 0.4$ (FWHM).  The goodness of fit worsens if the 
resolution is $\Delta log E = \pm 0.2$ (long dashes) or zero (short 
dashes).  In these two latter cases, the fits use $|m_{\nu}|= 0.25 
eV/c^2,$ and elemental abundances: 65\% A=1, 10\% A=4, 5\% A= 5 to 19, 
5\% A = 20 to 40, and 15\% A = 41 to 90.

No decent fits exist for $|m| \geq 0.75 eV/c^2$. All three fits would also
dramatically worsen if there were no near sources -- since the curves 
would then drop sharply at $E\approx 10^{19} eV.$ Thus, the flux beyond 
this energy appears in the model to come primarily from the 13 \% of 
sources that are``near."  A convenient way to represent changes in the 
composition of the CR's is to plot $<\ln A>$ versus energy -- see Fig.
2.  The model results are in rough agreement with the data in its 
essential features:  a rise of $<ln A>$ from the knee of the spectrum to 
a maximum near $10^{17}$ to $10^{18} eV$ and a subsequent decline to a 
near zero value, i.e., almost pure protons, at $10^{19} eV$. Given the 
difficulty in measuring the composition of CR's above the knee, such 
rough agreement is not unreasonable.

Exactly what is needed in the model to reproduce the specific features 
seen in the data?

{\bf The knee at $E\approx 4\times 10^{15} eV$} requires $|m_\nu|
\approx 0.5 eV/c^2$, so that threshold energy for CR protons to decay 
occurs at this energy, and the proton component drops
precipitously -- jagged curve in Fig. 1.

{\bf The $E^{-3}$ power law between $E= 10^{16}$ and $10^{18} eV$}
(near horizontal slope in Fig. 1) is reproduced only with
the choice of composition noted previously, and a large enough energy 
resolution to smooth out the bumps from different element thresholds.

{\bf The position of the dip at $E\approx 10^{19} eV$} also depends on
the $|m_\nu|$ value.  It occurs because at this
energy the threshold for the heaviest elements to decay is
reached, and the spectrum becomes depleted.

{\bf The rise for $E > 10^{19} eV$}, occurs because as $E$ increases, an
increasing fraction of A=1 particles from the near sources can reach us, 
given their lengthened lifetime and mfp.  This rise needs 13 \% of 
sources to be ``near," which is how the model ``explains" the apparent 
lack of GZK cutoff.\cite{Greisen,Takeda}

{\bf Composition vs energy (Fig. 2)} The composition is heavy
before the dip at $E\approx 10^{19}$ because only the heaviest elements 
are left in the spectrum at this $E$, since their thresholds have not
yet been reached.  However, at the highest energies the CR's are found 
to be very light, because by $E\approx 10^{19} eV$ the thresholds for 
all $A > 1$ nuclei have been reached, while this $E$ is far enough
above $E_{th}$ for A = 1 that this component is coming back.

{\bf The source spectrum} was chosen as $E^{-2.67}$ to match the 
observed spectrum below the knee.  Equivalently, any other power law 
$E^{-2.67+\alpha}$ could have been used if the effect of energy loss 
processes not included here were simply to steepen the source power law 
by $\alpha.$  (Of course, the dominant (A = 1) spectral component should
show very little energy loss due to other processes if the nucleons are 
neutrons during most of their time en route.)

\begin{figure}[htb]
\begin{center}
\leavevmode
\epsfxsize=3.25in
\epsffile{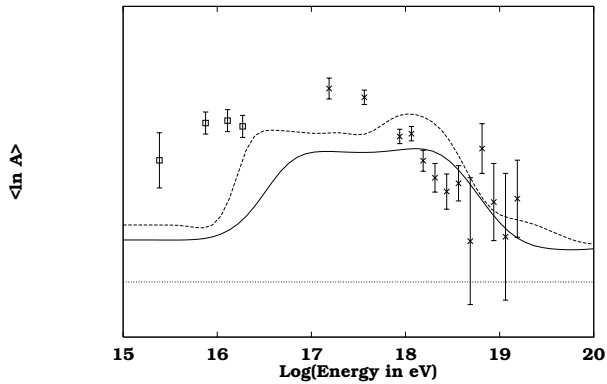}
\caption{Prediction of the model for the CR composition ($<ln
A>$) as a function of particle energy.  Solid and dashed curves makes 
the same assumptions for $|m_\nu|$, composition, and the
percentage of ``far" CR sources as the solid and dashed curves
shown in Fig. 1.  Data points with squares are BASJE (1994), and crosses 
are Fly's Eye (1993).}
\end{center}
\end{figure}

\section{Potential problems with model}

While the model may be consistent with some features of
the CR spectrum, that is a far cry from being evidence for
tachyonic neutrinos.  Let us consider a few of the problems with
the model.

{\bf Conventional explanations exist} for some of the regularities
we have noted, and plausible mechanisms exist to account for the 
production of the component of the spectrum believed to be galactic in 
origin.  However, few conventional explanations predict numerical values 
for the position of the knee and ankle, and many of the models have both 
ad hoc elements and free parameters.  Moreover, explaining some of the
spectral features represents a very severe test of all conventional
models -- particularly the {\it abruptness} of the change in slope at 
the knee and ankle.\cite{Erlykin}

{\bf A source composition independent of energy is highly unrealistic.}
But by making this assumption we are merely limiting the number of free 
parameters.

{\bf Other models can account for the absence of a GZK cutoff.} Various
suggestions have been made to explain why CR's with energies above the 
conjectured GZK cutoff ($E\approx 4\times 10^{19} eV$) apparently fail 
to be significantly degraded in energy by interaction with the 
CBR.\cite{Farrar,Coleman}  Nevertheless, as long as no specific distant 
sources have been identified, it would seem that the least exotic 
hypothesis is that CR sources with $E > 4\times 10^{19} eV$ simply are 
closer than a few dozen Mpc (as our model requires), {\it even if no 
specific sources have so far been identified.}

{\bf No mechanisms are known that have a single power law spanning
over ten decades.}  Of course, there are no known sources in the 
conventional theory of CR's at the highest energies either, though 
topological defects have been suggested as one possibility.\cite{Bhatt}  
But they have not been proposed to account for the lower energy region, 
which are believed to originate from supernova shocks.  One exotic 
possibility for sources has been proposed by Kuz'min and Tkachev: the
decay of supermassive long-lived particles produced in the early 
universe.\cite{Kuzmin}  One advantage of this possibility from our point 
of view is that such sources could be a considerable fraction of cold 
dark matter, and hence could be prominent in the Milky Way galactic 
halo, and therefore relatively nearby.  Yet, they would also be 
relatively isotropic, as seems to be the case for the limited number of 
events so far seen at the highest energies.

\section{Possible Confirming Tests}

The seven tritium beta decay experiments used by the Particle Data 
Group\cite{PDG} all report ${{m_\nu}_e}^2 < 0$.  Two of these
experiments report ${{m_\nu}_e}^2 < 0$ by over four standard
deviations $(4\sigma)$, but they are also $4\sigma$ 
apart.  Regrettably, the value we have used here
$|{{m_\nu}_e}|\approx 0.5 eV/c^2$ is too small to be consistent with
either of these experiments.  Moreover, the
tritium results have been explained in terms of either experimental 
anomalies,\cite{Barth,Lobashev} final state interactions, or new 
physics\cite{Stephenson} -- though some have attributed them to 
tachyonic neutrinos.\cite{Rembielinski,Ciborowski} If the electron 
neutrino really were a tachyon, could future tritium beta decay 
experiments test for values of ${{m_\nu}_e}^2 \approx 0.25
eV^2/c^4$?  The current systematic and statistical errors on $m^2$ are 
over an order of magnitude larger, so probably not without new types of
instruments.

If neutrinos really were tachyons, why should one put any more faith in
the mass obtained from a fit to the CR spectrum than the much larger 
values found in tritium experiments?  One answer is that the only 
statistically significant negative values found in tritium experiments 
are inconsistent, and have been attributed to other causes.
Secondly, if any of the masses from tritium experiments represented real 
tachyons, then the knee of the CR spectrum would have to occur one or 
two decades lower in energy than is observed, because the threshold 
energy for proton decay varies inversely with $|{{m_\nu}_e}|.$
Alternatively, if $|{{m_\nu}_e}|$ found from the CR spectrum fit is 
correct that only means that the values reported in the tritium 
experiments arise from causes other than tachyons.

Are there other places one might look for confirmation of the tachyonic
neutrino hypothesis?  Neutrino oscillation experiments, being sensitive 
to $\Delta m^2$ cannot reveal whether individual neutrino flavors have 
$m^2 < 0,$ and mass limits from the 1987A or future supernovae would
seem to lack the needed sensitivity.  There is, however, one unambiguous
test of the tachyonic neutrino hypothesis involving a CR neutron flux -- 
see fig. 1.


\begin{figure}[htb]
\begin{center}
\leavevmode
\epsfxsize=3.25in
\epsffile{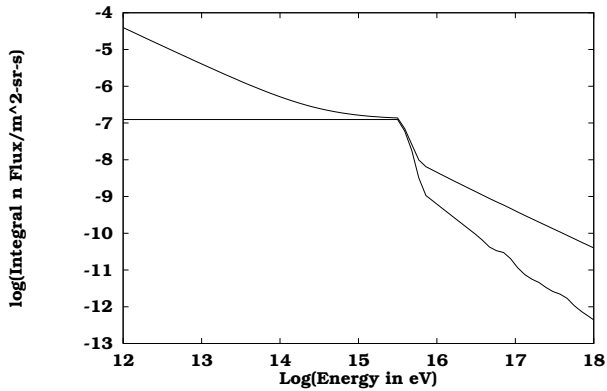}
\caption{Lower curve shows the prediction of the model for the log of 
the CR flux of neutrons integrated above an energy E, assuming $|m_\nu| 
= 0.5 eV/c^2$, 13\% near sources, with no convolution to account for 
finite energy resolution.  The upper curve shows the log of the 
integrated neutron flux atop a hypothetical $1/E$ background one tenth 
its amplitude at the position of the spike.}

\end{center}
\end{figure}

The signature of the model is a spike of neutrons just above the threshold
energy for proton beta decay at $E = 4.5\pm 2.2\times 10^{15} eV.$  The
uncertainty in the spike's position corresponds to the range: $0.25< |m| 
<0.75 eV/c^2$  The pile
up of neutrons just above $E_{th}$ is a consequence of the fractional 
energy loss of the nucleon becoming very small as $E_{th}$ is approached 
from above. Given distances to CR sources, virtually all neutrons below 
$E_{th}$ decay to protons long before reaching Earth.  As can be seen in 
fig. 3, the neutron spike might even be seen in plots of the integrated 
flux if the background were small enough.

Based on air shower measurements, it may be impossible to distinguish 
individual n's from p's in the region of the knee of
the spectrum, But, there is one clear difference: unlike protons or 
nuclei, multiple neutrons should point back to specific sources.  
Moreover, given the neutron lifetime, the mfp before decay at an energy 
of $10^{16} eV$ is only about 200 ly -- much too close for many sources
in any conventional model.  As Fig. 1 shows, neutrons should also be
seen as a large component of the flux at energies above $10^{19} eV.$  
However, if neutrons were seen at these energies, they could well be the 
result of sources closer than 0.2 Mly, and they would, therefore, have
little value in confirming the hypothesis of tachyonic neutrinos.



\acknowledgements

The author thanks Peter Becker, Jochen Bonn,
Alan Chodos, Robert Ellsworth, Alan Kosteleck\'{y},  Dietrich Muller, 
Jonathan Ormes, Len Ozernoy, Jakub Rembielinski, Todor Stanev and John 
Wilkerson for helpful suggestions.

\end{document}